\documentclass[preprint2]{aastex}

\shortauthors{Orosz \& Wade}
\shorttitle{Accretion Disks with Non-Steady
$T(r)$ Laws}

\begin{document}
\title{Ultraviolet Spectra of CV Accretion Disks with Non-Steady
$T(r)$ Laws}

\author{Jerome A. Orosz} 
\affil{Department of Astronomy, San Diego State University,
5500 Campanile Drive, San Diego, CA 92182}
\email{orosz@sciences.sdsu.edu}

\and

\author{Richard A. Wade}
\affil{{Department of Astronomy \& Astrophysics, 
The Pennsylvania State University, 525 Davey Laboratory, University Park,
PA  16802-6305}
\email{wade@astro.psu.edu}}
 

\begin{abstract} 
An extensive grid of synthetic mid- and far-ultraviolet spectra for
accretion disks in cataclysmic variables has been presented by Wade
and Hubeny (1998). In those models, the disk was
assumed to be in steady-state, that is $T_{\rm eff}(r)$ is specified
completely by the mass $M_{\rm WD}$ and radius $R_{\rm WD}$ of the
accreting white dwarf star and the mass transfer rate $\dot{M}$ which
is constant throughout the disk.  In these models, $T_{\rm
eff}(r)\propto r^{-3/4}$ except as modified by a cutoff term near the
white dwarf.

Actual disks may vary from the steady-state prescription for $T_{\rm
eff}(r)$, however, e.g.\ owing to outburst cycles in dwarf novae
($\dot{M}$ not constant with radius) or irradiation (in which case
$T_{\rm eff}$ in the outer disk is raised above $T_{\rm steady}$).  To
show how the spectra of such disks might differ from the steady case,
we present a study of the ultraviolet (UV) spectra of models in which
power-law temperature profiles $T_{\rm eff}(r) \propto r^{-\gamma}$
with $\gamma < 3/4$ are specified.  Otherwise, the construction of the
models is the same as in the Wade \& Hubeny grid, to allow comparison.
We discuss both the UV spectral energy distributions and the
appearance of the UV line spectra.  We also briefly discuss the
eclipse light curves of the non-standard models.
Comparison of these models with UV observations of novalike
variables suggests that better agreement may be possible
with such modified $T_{\rm eff}(r)$ profiles.
\end{abstract}


\section{Introduction}\label{intro}


In most types of cataclysmic variable (CV) stars matter accretes onto
the white dwarf via an accretion disk.  In some classes of CVs,
e.g.\ the novalike systems or dwarf novae in outburst, the accretion disk
can dominate the flux in the ultraviolet (UV) part of the spectrum. 
Several studies of the UV spectra of
uneclipsed CV disks have shown that we do not have a confident
understanding of how the photospheric spectrum is formed.
The disks in novalike variables (such as the prototype UX UMa
and also V603 Aql and RW Sex) should be
close to ``steady-state'' where
$$
T_{\rm eff}(r)=T_{\rm ref}\left({r\over R_{\rm WD}}\right)^{-3/4}
\left[1-\left({r\over R_{\rm WD}}\right)^{-1/2}\right]^{1/4}
$$
where $r$ is distance from the white dwarf and $R_{\rm WD}$
is the radius of the white dwarf.  The reference temperature
$T_{\rm ref}$ depends on the mass of the white dwarf $M_{\rm WD}$, its
radius, and the mass accretion rate $\dot{M}$:
$$
T_{\rm ref}
=\left[{3GM_{\rm WD}\dot{M}\over 
8\pi\sigma R_{\rm WD}^3}\right]^{1/4}
$$
where $G$ and $\sigma$ are the usual physical constants (Pringle 1981).

Modelling attempts that take into account the detailed vertical
structure and local emitted spectra from the disk surface have so far
shown considerable discrepancies between predicted and observed energy
distributions or $T(r)$ structures, especially at short wavelength
(e.g.\ Wade 1988; Long et al.\ 1994; Knigge et al.\ 1997; Robinson,
Wood, \& Wade 1999).  A general feature of the failing of these models
is that they predict ultraviolet (UV) spectra that are too ``blue''
compared with the observations.

To try to improve agreement between models and observation, the computation
of the local emitted spectra can be made more sophisticated in various
ways. Alternatively, additional source of UV radiation have been
proposed to be important in some cases (e.g. optically thin emission
from the base of a disk wind, Knigge et al.\ 1998).  However, it is
also worth considering whether the standard $T_{\rm eff}(r)$ formula
given above should be modified.  This may be because the disk is
non-stationary (e.g., a dwarf nova near outburst maximum may not have
relaxed to a steady or quasi-steady state), or because additional
local heating besides the viscous dissipation inside the disk may
alter the energy radiated locally at the disk surface (an example
would be strong irradiation of the outer disk by the central star, as
considered by, e.g., Vrtilek et al.\ 1990 in the context of low-mass
X-ray binaries), or for various other reasons.

In attempting to unravel the true structure of accretion disks,
studies of the emergent line spectrum should reveal much, because each
absorption line is formed in a relatively narrow range of temperatures
(corresponding to a narrow range of radii within the disk), and thus
has a characteristic Doppler width from projected disk rotation at the
characteristic radius.  (See Wade, Diaz \& Hubeny 1996 for
illustrative examples.)  An aspect of any study that considers an
altered $T_{\rm eff}(r)$ profile should therefore be to consider its
effect on the line spectrum from the disk.

\begin{figure*}[t]
\centerline{\includegraphics[width=4.3in,scale=1,angle=-90]{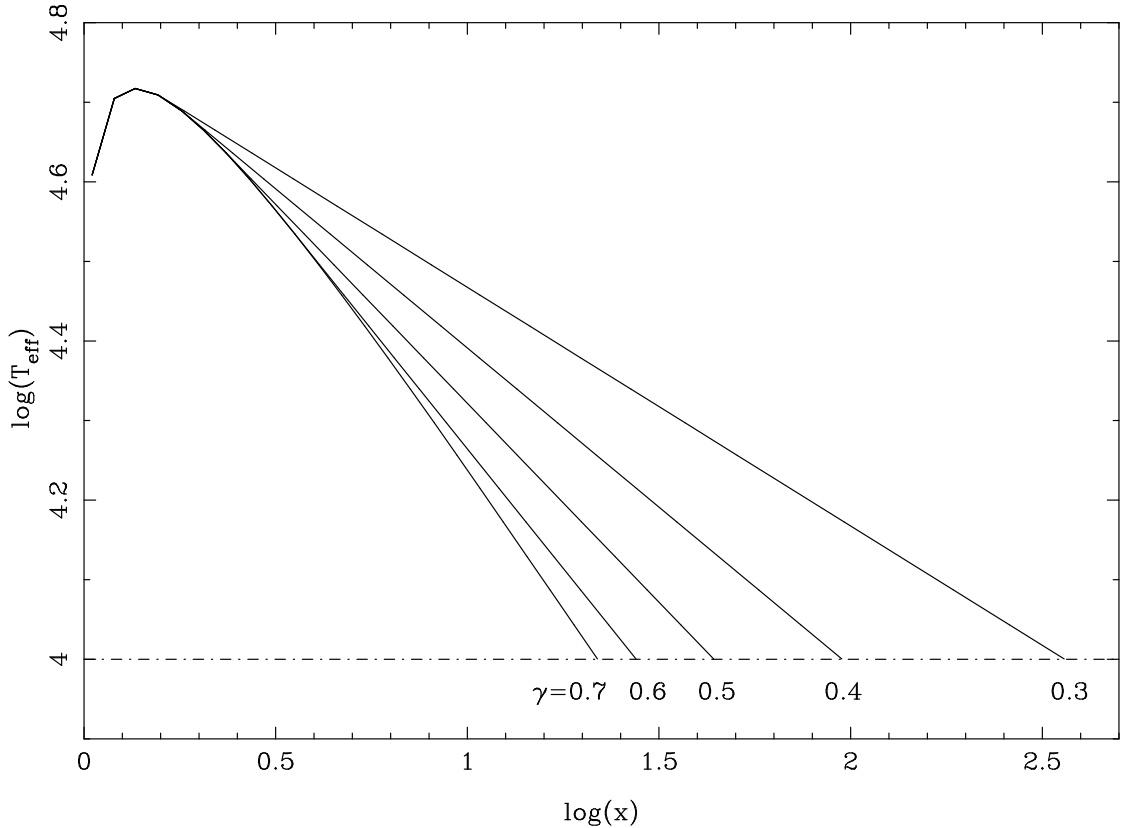}}
\figcaption[f1.eps]{The 
temperature profiles for a class of non-standard disks based on the
the WH98 model $v$ are shown.  Here $x$ is the dimensionless radius
defined by $x\equiv r/R_{\rm WD}$.  The segmented appearance of
the graph at small radii reflects the discrete set of radii used in
constructing the models and is not a physical feature.
\label{fig1}}
\end{figure*}

Wade \& Hubeny (1998, hereafter WH98) presented a grid
of detailed models for accretion disks in CVs. The WH98 models are
computed self-consistently in the plane-parallel approximation,
assuming LTE and vertical hydrostatic equilibrium, by solving
simultaneously the radiative transfer, hydrostatic equilibrium, and
energy balance equations.  Line transitions of elements from H to Ni
are accounted for in the computation of the local spectra of
individual disk annuli. 
Limb-darkening (Diaz, Wade, \& Hubeny 1996), Doppler broadening, and
blending of lines are fully accounted for in the computation of the
integrated disk spectra.

In this work we construct accretion disk models for cataclysmic
variables (based on the WH98 grid and its component local spectra), in
which the temperature profiles $T_{\rm eff}(r)$ are allowed to deviate
in a prescribed way
from the standard prescription\footnote{In what follows, we usually
suppress the subscript on $T_{\rm eff}$, but by $T(r)$ we always mean
the effective temperature.}.  We consider profiles that take a
power-law form $T_{\rm eff}(r)\propto r^{-\gamma}$ in the outer disk,
with $\gamma< 3/4$.  This does not exhaust the possibilities of
modified disks, but the choice is motivated by simplicity, 
and the intriguing and puzzling indication
(explored below) that such disks may in some
sense better describe the observations of novalike variables.

We discuss in the sections below our computational method and results,
including the UV spectral energy distributions, the appearance of the
UV line spectra, and briefly the UV light curves of eclipsing systems.

\section{Method}\label{method}

The computation of a WH98 model spectrum involves four stages, which
we summarize briefly (see WH98 for the detailed discussion).  First,
the disk is divided up into concentric annuli, which are treated as
independent plane-parallel radiating slabs.  The vertical structure of
each ring is computed with the program TLUSDISK (Hubeny 1990a, 1990b).
Second, the program SYNSPEC (Hubeny, Lanz, \& Jeffrey 1994) is used to
solve the radiative transfer equation to compute the local rest-frame
spectrum for each ring.  This computation includes typically thousands
of lines selected from the lists of Kurucz (Kurucz 1991; Kurucz \&
Bell 1995).  Third, the program DISKSYN5a is used to combine the rest
frame intensities and generate the integrated disk spectrum.
DISKSYN5a divides each ring into a large number of azimuthal sectors,
and sums the individual elements with the appropriate area weighting
and Doppler shifting (owing to the projected orbital motion of the gas
within the disk).  Finally, the program ROTINS is used to convolve the
integrated disk spectrum with a Gaussian instrumental broadening
function and re-sample the result uniformly in wavelength.

In the present study, 
modified temperature profiles are generated that have $T(r)\propto
r^{-\gamma}$ in the outer disk, and that match smoothly onto the
steady-state $T(r)$ profile at a tangent point.  The modified models
therefore share the standard model's $T(r)$ behavior in the inner disk
and depart smoothly from it further out.  The dimensionless radius of
the tangent point, $x_{\rm crit}\equiv r/R_{\rm WD}$, is given by
$$ x_{\rm crit} = \Bigg({7-8\gamma \over 6-8\gamma}\Bigg)^2$$
(valid for $\gamma < 0.75$).  Table \ref{tab1} 
lists $x_{\rm crit}$ and other
useful quantities for a selection of values of $\gamma$.  Figure 1
illustrates a family of modified $T(r)$ profiles for a particular
model ($v$) from WH98.  Note that $\gamma = 0.75$ is just the
steady-state case (as $x_{\rm crit}\rightarrow \infty$).
We emphasize that our particular class of models is chosen because it
allows a convenient parametrization for exploratory purposes that has
the standard model as a limiting case. Other classes of
models can be contemplated (e.g., a single power-law behavior of
$T(r)$ throughout the entire disk) that would give qualitatively
similar results, as well as model types that might give considerably
different behavior.  Since observed disks seems to differ from the
standard model disks by being ``redder'', our considered class of
models is possibly representative of the character of plausible
changes in T(r) structure.  

We consider modifications of the standard
(i.e., steady-state) WH98 models that have $M_{\rm
WD}=1.03\,{\rm M_{\odot}}$, namely models $t$, $u$, $v$, $dd$, and $jj$.
The radius of the central white dwarf for these models is $R_{\rm WD}
= 5.18\times 10^8$~cm; this is also taken to be the inner edge
of the accretion disk.  For discussion of eclipse light curves in the
UV, we make use of models based on WH98 models $k$ and $bb$, which
have $M_{\rm WD}=0.55\,M_{\odot}$ and $R_{\rm WD} = 9.05\times
10^8$~cm.  Table \ref{tab2} 
summarizes some properties of these models.  For
each of these standard WH98 models we generated modified models for
$\gamma=0.7$, 0.6, 0.5, 0.4, and 0.3.

The construction of the final model spectra is the same as for the
WH98 grid, except for the modified temperature profile of the disk and
the specification of the outer radius of the disk.  We prescribe the
outer edge of the disk in two ways.
In the first case the outer edge of the disk is where $T_{\rm eff}$
falls below about 10,000~K, as in WH98.  Since we consider only the
far- and mid-ultraviolet fluxes, the restriction to temperatures
greater than about 10,000~K results in flux errors of less than a few
percent.
In the second case, the outer edge of the disk is defined by the tidal
truncation radius appropriate to cataclysmic variables.  If $M_{\rm
WD}=1.03\,{\rm M_{\odot}}$, the outer disk edge would be at about
$50\,R_{\rm WD}$ for an orbital period of $P=1.5$ hours or at about
$100\,R_{\rm WD}$ for an orbital period of $P=6.0$ hours.  (These
limits were computed using Roche geometry and $M_{\rm sec} = 0.1 P$,
where $P$ is measured in hours and $M_{\rm sec}$ is the mass of the
secondary star in solar units.)  Table \ref{tab2} shows that truncation of the
disk at a temperature well above 10,000~K is likely in the case of
high mass transfer rates or low values of $\gamma$.

Once the modified radii are computed, the integrated disk spectrum
is computed in exactly the same way as the standard WH98 models:
local spectra are generated and summed together with appropriate area
weighting, taking into account Doppler shifts (owing to orbital motions
of the area elements within the disk) and limb darkening.  
We chose inclination angles of $8^{\circ}$,
$60^{\circ}$, and $76^{\circ}$ ($\cos i=0.99$, 0.50, and 0.25)
for illustration.
The fluxes presented here are given for an assumed distance
of 100 pc, and correctly account for limb darkening, but
do {\em not} include a geometrical foreshortening
factor $\cos i$.

Note that in these exploratory calculations, we have deviated from
requiring that the local $T_{\rm eff}$ and photospheric gravity $\log
g$ be mutually consistent: rather than compute new vertical structures
and local spectra for each annulus, we have simply used the vertical
structures and spectra from WH98, shifting each model annulus to a new
radius according to its $T_{\rm eff}$.  The main results of this study
of the ultraviolet spectrum should not be greatly affected by this shortcut.

\section{Results}\label{results}

\subsection{``Color-Magnitude'' Diagrams}

\begin{figure*}[ht]
\centerline{\includegraphics[width=4.1in,scale=1.0,angle=-90]{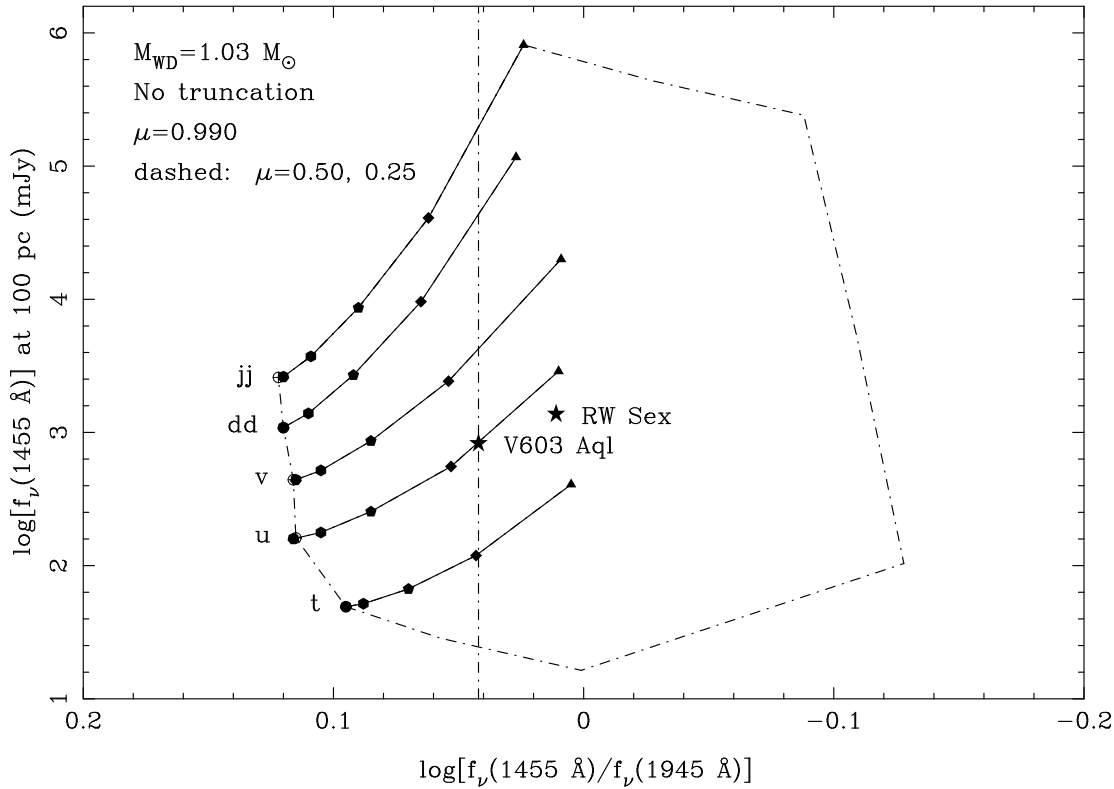}}
\figcaption[f2.eps]{A 
``color-magnitude'' diagram for the integrated disk spectra
where the ``color'' is
defined as $\log (f_{\nu}(1455 {\rm \AA})/f_{\nu}(1945 {\rm \AA}))$
and the ``magnitude'' is defined as $\log f_{\nu}(1455 {\rm \AA})$
at 100 pc, with $f_{\nu}$ in mJy.  The disks are viewed nearly
face-on ($\cos i=0.99$) and are allowed to extend outwards until
$T_{\rm eff}\approx 10,000$~K.  The solid lines connect sequences
of models with the same maximum temperature $T_{\rm max}$ and
different $\gamma$.  
The dashed box outlines the region occupied by models
of disks inclined to the line of sight,
with limb darkening taken into account but geometrical foreshortening
ignored (see text).  
A spectral index $\alpha=1/3$, where $\alpha$ is defined by
$f_{\nu}\propto \nu^{-\alpha}$, is shown by the vertical dashed line.
Points representing {\em IUE} observations of the novalike variables
RW Sex and V603 Aql are also shown (see text for details).
\label{fig2}}
\end{figure*}

\begin{figure*}
\centerline{\includegraphics[width=3.8in,scale=1,angle=-90]{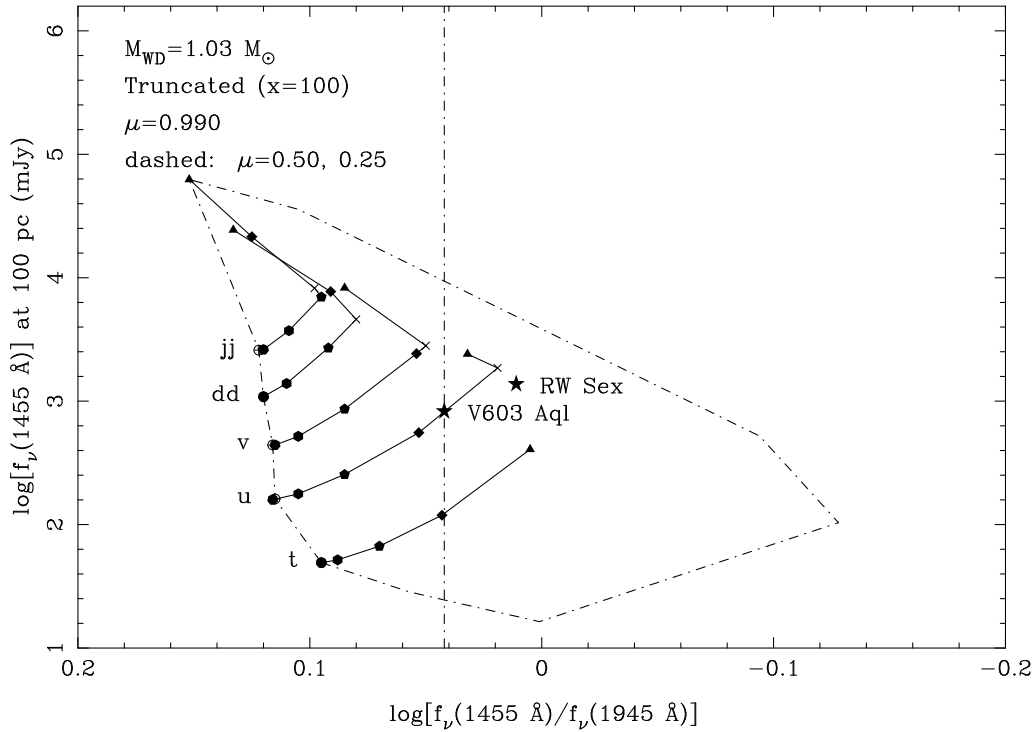}}
\figcaption[f3.eps]{Similar 
to Figure \protect\ref{fig2}, except that the
disks have been truncated at a dimensionless radius of
$x=100$.
\label{fig3}}
\end{figure*}

We show ``color-magnitude'' diagrams of the resulting disk spectra in
Figures \ref{fig2} and \ref{fig3}.  We use near-monochromatic fluxes
that are averages of $f_{\nu}$ over 10~\AA\ intervals centered at
1455~\AA\ and 1945~\AA.  These intervals are relatively line-free.
Note that the flux ratio $f_{\nu}(1455~{\rm \AA})/f_{\nu}(1945~{\rm \AA})$
is very nearly independent of interstellar reddening (Seaton 1979).
Figure \ref{fig2} shows power-law disks that are viewed nearly face-on
($\cos i=0.99$) and are not truncated (i.e.\ the disks extend until an
effective temperature of about 10,000~K is reached).  
The solid lines connect sequences of models with 
the same maximum temperature and different 
$\gamma$.
For $\gamma < 0.75$, the disks are hotter at each radius in the outer
disk than in the corresponding standard case, the disks extend further
out in radius, and there is a larger proportion of the disk surface at
relatively low temperatures. These disks are brighter and {\em
redder}, that is, $\log f_{\nu}(1455 {\rm \AA})/f_{\nu}(1945 {\rm
\AA})$ is more negative, displayed further to the right in Figure
\ref{fig2}.
The dashed box outlines the region occupied by models when limb
darkening is taken into account\footnote{To make the meaning
of this region more explicit, consider the point marked ``t''
in Figure \ref{fig2}.  This is model {\em t} in the case of the
standard disk and for $\mu = 0.990$. The edge of the dashed
box descending to the right from this point is the track of
this model as $\mu$ decreases to 0.250.  The next vertex ascending
to the right is model {\em t} with $\gamma = 0.3$ and $\mu = 0.250$.
The vertex descending to the right from the point representing
model {\em jj} (with $\gamma=0.3$ and $\mu = 0.990$) 
is model {\em jj} with $\gamma = 0.3$ and $\mu = 0.250$.  For 
all the other models explicitly shown (with $\mu = 0.990$), the
corresponding points for $\mu = 0.500$ and 0.250 lie within the
dashed box, descending to the right from the $\mu= 0.990$ case.
Figure \ref{fig3}\ is similar, with the dashed vertex at
coordinates $(-0.094, 2.717)$ representing the critical case for model
{\em u} for $\mu = 0.250$, and the vertex at $(0.105, 4.554)$
representing the truncated model {\em jj} for $\mu = 0.250$.
}.  
(Models with $\cos i=0.50$ and 0.25
are included in defining this box. As before, the calculated fluxes
are scaled to a distance of $d=100$ pc. The purely geometrical
foreshortening factor $\cos i$ is {\em not} included.)  At high
inclinations, limb darkening lowers the disk surface brightness and
reddens the disk.  ``Cooler'' models such as model $t$ and models with
$\gamma=0.3$ have more pronounced limb darkening effects in the
ultraviolet (see Diaz, Wade, \& Hubeny [1996] for further discussion of 
limb darkening in accretion disk spectra.)

Figure \ref{fig3} shows what happens when the model disks are
truncated at a dimensionless radius of $x=100$.  The dashed box again
outlines the region when high inclination angles are considered.  In
this case a disk becomes brighter and redder as $\gamma$ declines from
0.75, as long as the entire disk with $T_{\rm eff}(x) > 10,000$~K fits
within $x=100$, as before.  For $\gamma$ less than some critical value
$\gamma_{\rm crit}$, however, the disk is no  longer
allowed to grow in radius, but
each part of the disk within $x=100$ continues to get hotter, so the
disk gets brighter but {\em bluer}.  At a fixed value of
$\gamma$, truncated disks are fainter and bluer than non-truncated
disks.

For truncation at $x=50$, corresponding to shorter orbital periods,
the result is that the disk is smaller, fainter, and bluer than the
corresponding disk truncated at $x=100$.  Such disks are still
brighter than the standard disks from which they were derived but may
be either redder or bluer depending on $\gamma$ and maximum
temperature $T_{\rm max}$.

\begin{figure*}[t]
\centerline{\includegraphics[width=4.2in,angle=-90,scale=1]{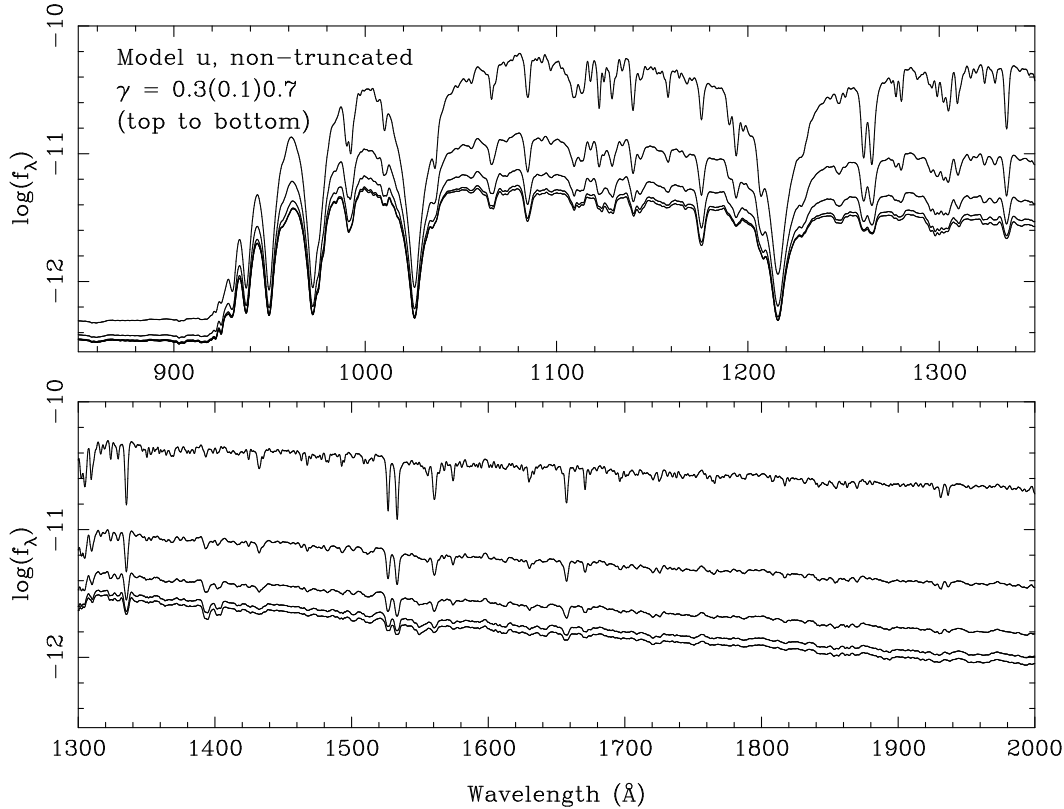}}
\figcaption[f4.eps]{The 
integrated spectra at $\cos i=0.99$ for the non-truncated
$\gamma$ disks derived from the WH98 model $u$ are shown.  The
fluxes are for a distance of $d=100$ pc.
\label{fig4}}
\end{figure*}

In both Figures \ref{fig2} and \ref{fig3}, points are plotted which
represent {\em IUE} observations of the novalike variables RW Sex and
V603 Aql (cf.\ Verbunt 1987).  The data have been corrected for
extinction and reddening and scaled to $d=100$ pc and $\cos i=0.99$
(i.e., geometrical foreshortening removed) for comparison with the
models.  For RW Sex, $d=150$~pc and $\cos i = 0.73$ are used; for V603
Aql, $d= 376$~pc and $\cos i = 0.96$, following Wade (1988).  The
orbital periods are 5.9 hours and 3.3 hours, respectively.  The disk
models for the standard cases are too blue to match the observed
colors of RW Sex and V603 Aql.  This point will be discussed further
below.  

\begin{figure*}[p]
\centerline{\includegraphics[width=4.9in,scale=1.2]{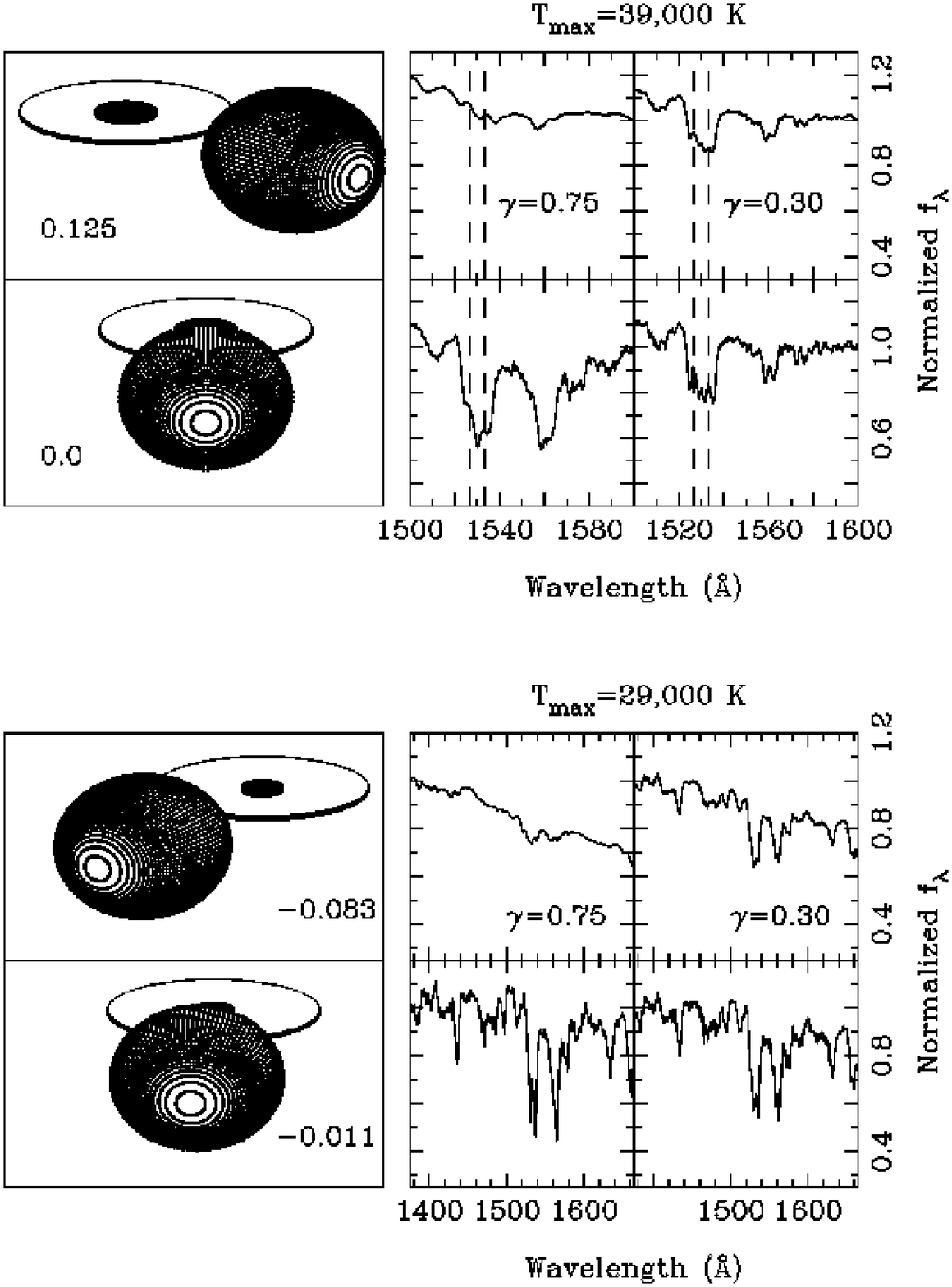}}
\figcaption[f5.eps]{Eclipse geometry and representative
spectra for for models $bb$ (top) and $k$ (bottom).  See text for
model parameters.
Left panels: The eclipse geometry, as seen by a distant observer.  The
orbital phase is indicated (mid--eclipse is zero phase).  The dark
inner regions of each disk represent the annuli of the steady-state
models $\gamma = 0.75)$ that have $T_{\rm eff}\gtrsim 10,000$~K.  The
outer rings represent the outer edge of the disks for the $\gamma=0.3$
models (in both cases $x\approx 51$).  The disk for model $k$ is not
truncated (i.e.\ the last ring has $T_{\rm eff}\approx 10,000$~K),
whereas the last ring for the $bb$ model with $\gamma=0.3$ has $T_{\rm
eff}\approx 13,500$~K.
Right panels: Integrated spectra at $R=25,000$ (top, $bb$ models) and
$R=2000$ (bottom, $k$ models) for the steady-state and $\gamma=0.3$
cases.  The dashed lines in the upper plots indicate the rest
wavelengths of the Si II doublet.
\label{fig5}}
\end{figure*}

WH98 discuss the contribution of the central white dwarf (WD) to the
light from the disk (their Section 4.3).  For the relatively hot disks
that are considered here, the WD's contribution does not make a
qualitative difference to the discussion, owing to its small surface
area in comparison with the outer disk, and has been omitted from our
exploratory calculations.  This is especially true for the
non-standard disks of the kind considered here, which are more
luminous than the standard disks.  For realistic cases, the WD
contribution to the models summarized in Figures \ref{fig2} and
\ref{fig3}\ would be less than $\log f_{\nu}(1455~{\rm \AA}) \approx$ 1 or
1.5 (mJy at 100 pc).  There will be a noticeable contribution to the
(out of eclipse) flux of a CV only when the WD is hotter than the
inner disk, and in those cases the total light will be somewhat bluer
than a standard disk; in the interesting non-standard cases, the WD
contribution is even less significant. Of course, the effect of the WD
might be seen during deep eclipses of the disk, and this should be
modeled in detailed studies of observed systems; this does not make a
qualitative or important quantitative change to the conclusions of
this study in any way.  Similar considerations apply to the possible
contribution from a boundary layer (BL) at the inner edge of the disk,
but with even more force because the BL radiating area would be
smaller than the WD surface area and the UV bolometric correction
would be very large, placing most of the emission in the extreme UV or
soft X-ray region.

\subsection{Integrated Spectra}

Figure \ref{fig4} shows integrated spectra at $\cos i=0.99$ for
non-truncated $\gamma$ disks derived from the standard WD98 model $u$.
The disks with flatter temperature distributions (i.e.\ smaller
$\gamma$) are brighter.  Fluxes are again for a distance of $d=100$ pc
and do {\em not} include the foreshortening factor $\cos i$ (which is
however the same for all spectra shown).  In addition to the overall
flux level and color changes, specific line features that are formed
in the outer (cooler) portions of the disk are enhanced relative to
features that are formed in the inner disk.  For example, compare the
C IV doublet at $\lambda\approx 1548, 1550$~\AA, which is formed in
the hotter regions of the disk, with the Si II doublet at
$\lambda\approx 1527, 1533$~\AA.  In the model spectrum with
$\gamma=0.7$, the Si II feature is only slightly stronger than the C
IV feature, whereas the Si II feature is noticeably stronger in the
model spectrum with $\gamma=0.3$.

\subsection{Extension to Eclipsing Systems}

Wade \& Orosz (1999) discuss the modifications to the WH98 models to
deal with the case of a disk that is partially eclipsed by the
mass-losing star.  We used the Wade \& Orosz codes to compute spectra
and ``light curves'' through various partial eclipse phases for
non-standard models derived from the WH98 models $bb$ ($T_{\rm
max}=39,000$~K) and $k$ ($T_{\rm max}=29,000$~K).
Both models have a white dwarf mass of $M_{\rm WD}=0.55\,{\rm
M_{\odot}}$.  For the illustrations we choose a secondary star mass of
$M_{\rm sec}=0.5\,{\rm M_{\odot}}$, an orbital period of 0.2 days, an
inclination of $i=71^{\circ}$, and an outer disk radius of $x\approx
51$.  We chose two different spectral resolutions at which to sample
the final model spectra: $R=25,000$ (FWHM = 0.062~\AA\ sampled every
0.0155~\AA) and $R=2000$ (FWHM = 0.78~\AA\ sampled every
0.195~\AA). The former resolving power was delivered by the G160M
grating of the Goddard High Resolution Spectrograph (GHRS) which flew
on the {\em Hubble Space Telescope}, and the latter resolving power was
delivered by the G140L GHRS grating.

\begin{figure*}[t]
\centerline{\includegraphics[width=4.1in,scale=1.0]{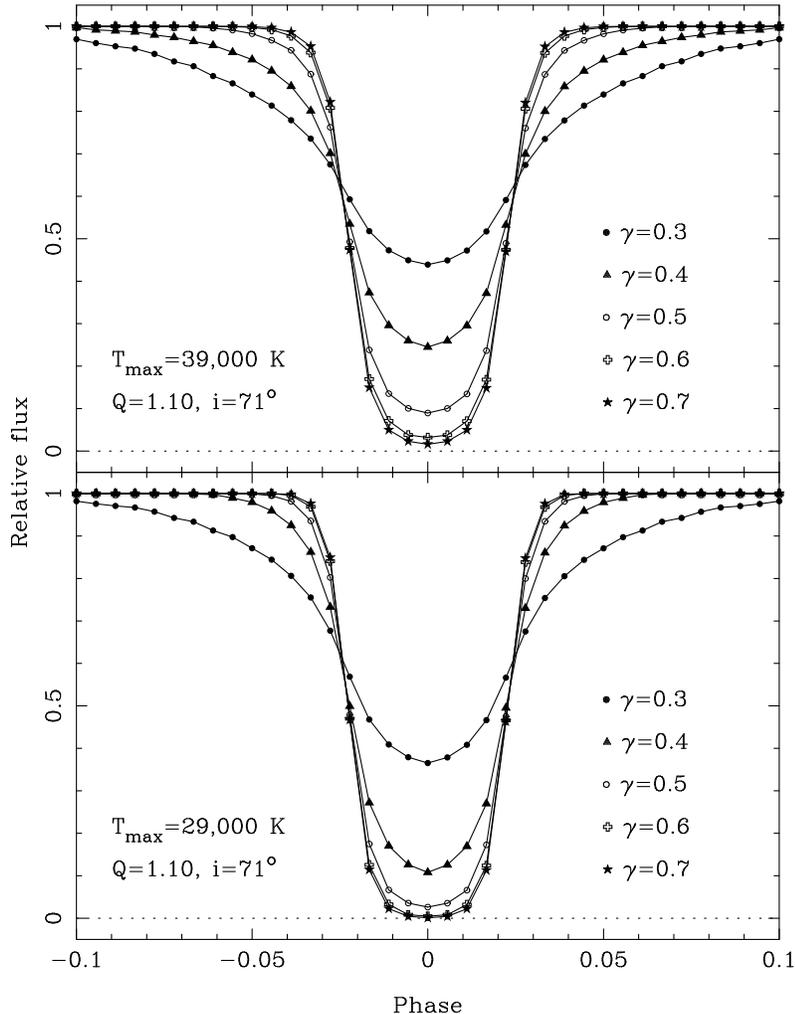}}
\figcaption[f6.eps]{Normalized 
eclipse light curves for various $\gamma$-models derived from 
models $bb$ (top) and $k$ (bottom). 
\label{fig6}}
\end{figure*}

Figure \ref{fig5} shows the eclipse geometry and predicted spectra (in
the restframe of the white dwarf) in the region of the \ion{Si}{2}\
doublet ($\lambda=1526.71$, 1533.45~\AA) for models $bb$ ($R=25,000$)
and $k$ ($R=2000$) at different phases, where mid-eclipse is phase 0.0
(for clarity each spectrum has been normalized at 1600~\AA).  We
display only the rings in the disk that have $T_{\rm eff}\gtrsim
10,000$~K.
The integrated disk spectra for the standard models ($\gamma=0.75$)
only show weak features outside of eclipse.  As the hotter inner
regions of the disk are eclipsed, lines that are formed in the outer
regions of the disk (for example \ion{Si}{2}) become apparent.  As
noted above, specific line features that are formed in the outer disk
are enhanced relative to features that are formed in the inner disk in
the models with flatter temperature profiles.  Hence, the \ion{Si}{2}
feature in the uneclipsed disks with $\gamma=0.3$ is relatively
strong.  Because the outer regions of the disk are relatively large in
the $\gamma=0.3$ models, the change in the line spectra of the
$\gamma=0.3$ models through eclipse is much less dramatic than in the
corresponding standard model cases.

Figure \ref{fig6} shows continuum light curves through eclipse for
various models based on $bb$ and $k$ (for a mass ratio $Q\equiv M_{\rm
WD}/M_{\rm sec} = 1.1$ and an inclination of $71^{\circ}$).  Here the
``continuum'' represents the average flux between 1410~\AA\ and
1530~\AA\ in the $R=2000$ spectra.  The depth of the light curve at
phase 0.0 depends on the value of $\gamma$.  The disk in the standard
model $k$ is totally eclipsed at phase 0.0 for our adopted geometry,
whereas the flux at phase 0.0 for the $\gamma=0.3$ model based on $k$
is roughly 35\% of the out-of-eclipse level.
Since the models with flatter temperature profiles have larger disks
than the corresponding standard model, the eclipse light curves for models
with low values of $\gamma$ are much broader than those for the
standard models.  Hence one may use the shape and depth of the eclipse
light curve to help distinguish between standard models and models with
much flatter temperature profiles.

\section{Discussion}\label{disc}

There are several ways to compare the detailed model spectra with UV
observations of disks in CVs.  With the correct model, one should be
able to match the overall flux, assuming the distance and inclination
of the object in question are well enough known.  One should also be able
to match the overall spectral shape (the ``color'') over some broad
wavelength range.  The predicted absorption line spectrum can be
compared with high-quality observations for nearly face-on systems or
for systems in partial eclipse (Wade \& Orosz 1999).  Finally, the
predicted light curves for eclipsing systems can be compared with
observations.  In particular, the width and depth of the eclipse
light curve are useful for distinguishing between various models.

As discussed by Wade (1988), it turns out that UV observations of
novalike variables are very difficult or impossible to fit using
steady-state disk models constructed from LTE stellar atmospheres.
That is, one cannot match both the observed flux
and color at the same time.  This is still the case using the more
self-consistent disk models of WH98.  
Indeed, this is obvious from Figures \ref{fig2} and \ref{fig3}
(where the open--circle symbols represent the steady--state
disk spectra, $\gamma = 0 .75$).  While
the standard WH98 models $dd$ and $v$ roughly reproduce the observed
fluxes of RW Sex and V603 Aql, they fail badly in matching the colors.
The present work suggests that ``non-steady state'' disks with a
power-law $T_{\rm eff}(r)$ distribution that is flatter than
$r^{-3/4}$ might fare better.  For example, V603 Aql is on the
sequence of power-law disks based on the WH98 model $u$ ($T_{\rm
max}=39,000$~K) with $0.3\lesssim\gamma\lesssim 0.4$.

There is one obvious problem associated with using non-standard
$T_{\rm eff}(r)$ profiles, 
either of the class described here or more generally,
to model the disks in the novalike
variables: what would be the reason that the steady-state $T(r)$ law is not
followed in these apparently steady-state disks?  It is probably not
reasonable to explain a non-standard temperature profile as being due
to a mass transfer rate that is constant in time but varies with
radius.  This would result in sources or sinks of mass density within
the disk, in contradiction to the apparent steady-state nature of
these stars.   (See, however, Knigge (1999), who discusses how a
strong wind from the {\em inner} disk may act as a sink of
mass, angular momentum, and energy.)
Hence the non-standard temperature profile must be ascribed to some
other cause.  One possible suspect is irradiation of the disk by a hot
central white dwarf or boundary layer.  Vrtilek et al.\ (1990)
showed how irradiation can lead under certain assumptions to 
$\gamma\approx 3/7$ in
low-mass X-ray binaries, but note that in that case radiation from the
central source (a neutron star) dominates the outer disk flux, since 
the potential well into which the disk matter is accreted is
very deep.  It is far from clear that irradiation in
CVs (where the central star is a white dwarf), acting alone,
will be strong enough to enforce such a flat temperature profile,
and we do not claim to have shown that irradiation is the cause
of a flattened $T(r)$ profile\footnote{Indeed, we do not even
make the claim, on the basis of the limited work described here,
that a flattened $T(r)$ profile is the basis of the discrepancy
between standard disk models and observations of luminous CVs.  
We have only demonstrated consistency of such modified disks with
observations of two test objects, in terms of UV colors and fluxes, 
along with pointing to other more sensitive tests involving 
spectra and eclipse light curves.}.
However, if irradiation is the cause of the non-standard temperature
profile, then our models are incomplete---we have merely assigned the
same local spectra (computed under the assumption of {\em no}
irradiation) to different radii in the disk.  A correct treatment of
irradiation would include its effect in the vertical structure of the
disk (cf.\ Hubeny 1990a) and the emerging spectrum would
differ from merely a sum of hotter ``normal'' spectra.  An
irradiated disk spectrum might even resemble to some extent a sum of
Planck spectra, which Wade (1988) showed did match colors and fluxes
of novalike variables in the ultraviolet spectral range.  Until
detailed spectrum models incorporating a correct description of
irradiation are developed, the ``problem of the novalike variables''
must be considered to still be unsolved.

\section{Summary}

We have presented a class of model spectra of accretion disks which have
non-standard power-law temperature profiles (i.e., $T_{\rm
eff}(r)\propto r^{-\gamma}$ with $\gamma<0.75$).  When $\gamma<0.75$,
the disks are brighter and redder than disks in the corresponding
standard case, assuming the disks with the flatter temperature
profiles extend outward until a temperature near 10,000~K is
reached.  Non-standard disks that are truncated as some smaller radius
are still brighter than the standard disks from which they were
derived but may be either redder or bluer depending on $\gamma$ and
the maximum temperature $T_{\rm max}$.  
For smaller values of $\gamma$ (flatter temperature profiles),
specific line features that are formed
in the outer disk are enhanced in the integrated spectrum relative to
features that are formed in the inner disk.  For eclipsing systems we
have illustrated how the eclipse light curves for these non-standard
disks are broader and less deep than the light curves for the
corresponding standard disks.  
Such patterns may be useful in diagnosing whether disks in CVs do in
fact have non-standard $T(r)$ laws, or whether the failure of standard
model disks to fit the data lies elsewhere.

Finally, we point out that detailed standard disk models ($T_{\rm
eff}(r)\propto r^{-3/4}$) of Wade \& Hubeny (1998) cannot
simultaneously match the observed UV fluxes and colors of the novalike
variables RW Sex and V603 Aql.  Wade (1988) reached a similar
conclusion, using disk spectra synthesized from stellar atmosphere
models.  
Our present exploratory work suggests that a disk with a much flatter
temperature profile, $T_{\rm eff}(r)\propto r^{-\gamma}$ with
$\gamma\approx 0.35$, provides a better match to both the flux and
color of V603 Aql.  However, the WH98 models as applied in this case
are incomplete; since no mechanism is identified that results in the
flatter temperature profile, we have simply constructed models as if
there were a mass transfer rate that varies with radius (an apparently
unphysical situation for the novalike variables).  If the flat
temperature profiles which seem to be favored by this exploratory
analysis are in fact due to irradiation
(although for energetic reasons this seems problematic), 
then further refinements to
the models should include a detailed and consistent treatment of
irradiation.

\acknowledgements

Support from NASA grant NAGW-3171 and Space Telescope Science
Institute grants GO-0661.01-A and AR-07991.01-96A is gratefully
acknowledged.  
We are indebted to Ivan Hubeny for the use of his disk
codes and his continuing friendly and helpful advice.
This work has made use of the SIMBAD database, operated by
CDS at Strasbourg, France, and NASA's Astrophysics Data System
Abstract Service.




\clearpage

\begin{deluxetable}{cccccc}
\tablecaption{Data for Power-Law $T(r)$ Disk Models
\label{tab1}}
\tablewidth{0pt}
\tablehead{
\colhead{$\gamma$} & 
\colhead{$x_{\rm crit}$} & 
\colhead{$T(x_{\rm crit})/T_{\rm max}$} &
\colhead{$T(x=50)/T_{\rm max}$} &
\colhead{$T(x=100)/T_{\rm max}$}} 
\startdata
0.7 & 12.250 & 0.288 & 0.108 & 0.066 \cr
0.6 & 3.361 & 0.678  & 0.134 & 0.088 \cr
0.5 & 2.250 & 0.848  & 0.180 & 0.127 \cr
0.4 & 1.842 & 0.929  & 0.248 & 0.188  \cr 
0.3 & 1.633 & 0.969  & 0.347 & 0.282  \cr
\enddata
\tablecomments{See text for definition of $x_{\rm crit}$.
The dimensionless radius is
$x \equiv r/R_{\rm WD}$, where $r$ is the radial coordinate in the disk.
$T_{\rm max}$ is the maximum effective temperature of a
steady-state disk, attained at $x=1.36$.}
\end{deluxetable}


\begin{deluxetable}{cccccc}
\tablecaption{Properties of Disk Models
\label{tab2}}
\tablewidth{0pt}
\tablehead{
\colhead{Model} &
\colhead{} &
\colhead{$T_{\rm max}$} &
\colhead{}  &
\colhead{$T(x=50)$\tablenotemark{a}} &
\colhead{$T(x=100)$\tablenotemark{a}} \\
\colhead{Family}&    
\colhead{$\log \dot{M}$} &
\colhead{(K)}  &
\colhead{$x(T=10,000 K)$} &
\colhead{(K)} &
\colhead{(K)}}
\startdata 
t   &         -10.0  &      29,330  &      9.40    &     10,180 &   (8270) \cr
u   &          -9.5  &      39,110  &     14.30    &     13,580 &   11,030 \cr
v   &          -9.0  &      52,160  &     21.70    &     18,100 &   14,710 \cr
dd  &          -8.5  &      69,560  &     28.80    &     24,140 &   19,620 \cr
jj  &          -8.0  &      92,760  &     43.70    &     32,200 &   26,160 \cr
    &               &           &              &           &   \cr
k   &          -9.0 &       29,330 &       9.40  &       10,180 &  (8270) \cr
bb  &          -8.5 &       39,110 &      14.30  &       13,580 &  11,030 \cr
\enddata
\tablecomments{For model sequences $t$, $u$, $v$, $dd$, and $jj$,
$M_{\rm WD} = 1.03 {\rm M}_\odot$ and $R_{\rm WD} = 5.18 \times 10^8$ cm.
For model sequences $k$ and $bb$,
$M_{\rm WD} = 0.55 {\rm M}_\odot$ and $R_{\rm WD} = 9.05 \times 10^8$ cm.
The steady-state mass transfer rate $\dot{M}$ is in units of solar masses
per year.  $x\equiv r/R_{\rm WD}$, where $r$ is the radial coordinate
in the disk.  $T_{\rm max} = T_{\rm eff}(x=1.36)$.
All temperatures are effective temperatures.
See WH88 for further details.}
\tablenotetext{a}{$T(x=50)$ and 
$T(x=100)$ are shown for the case $\gamma = 0.3$.
Models discussed in the text are extended to $T_{\rm eff}= 10,000$~K
or are truncated at $x=50$ or $x=100$, if $T_{\rm eff}=10,000$~K has
not been reached at that radius.}
\end{deluxetable}

\end{document}